# Relativistic QED Plasma at Extremely High Temperature

Samina S. Masood[1]

Department of Physical and Applied Sciences, University of Houston
Clear Lake, Houston TX 77058

## ABSTRACT

Renormalization scheme of QED (Quantum Electrodynamics) at high temperatures is used to calculate the effective parameters of relativistic plasma in the early universe. Renormalization constants of QED play role of effective parameters of the theory and can be used to determine the collective behavior of the medium. We explicitly show that the dielectric constant, magnetic reluctivity, Debye length and the plasma frequency depend on temperature in the early universe. Propagation speed, refractive index, plasma frequency and Debye shielding length of a QED plasma are computed at extremely high temperatures in the early universe. We also found the favorable conditions for the relativistic plasma from this calculations.

## 1. INTRODUCTION

Renormalization is a process of removal of singularities in gauge theories. Perturbation theory is needed to calculate the radiative corrections to higher order processes in a medium. KLN (Kinoshita-Lee and Nauenberg) theorem [1, 2] requires an order by order cancellation of singularities to assure the finiteness of a theory at all orders of perturbative expansion and works perfectly for QED in vacuum. The coupling constant of the theory, as an expansion parameter of QED perturbative series, is sufficiently small to suppress the higher order effects in vacuum and ensures the finiteness of QED parameters. However, this coupling constant starts to increase at temperatures larger than the electron mass and a relativistic QED plasma can exist.

Renormalization of gauge theories at extremely high temperatures needs thermal modification of particle propagators of the theory. This modified propagators may be identified as thermally dressed propagators and each one of them has an additional term in real-time formalism [3-4] to incorporate the probability of interaction of propagating particle with the medium. The interaction with the medium induce temperature dependence to the physically measurable parameters [5-13] of the theory such as electron mass, charge and wavefunction in the system and these parameters quadratically grow with temperature. Impact of this temperature dependence on astrophysical systems [14-15] is already known.

It has been previously shown that the temperature affects the electromagnetic properties of the medium itself which determine the behavior of the system under extreme conditions [3-15]. Renormalization constants of QED in a medium refer to physically measureable parameters of the theory which affect the electromagnetic properties of the medium and are associated with dynamically generated plasma screening mass [3,7,13, 16] of photon and the propagation speed of photons. Debye shielding length is also calculated as a function of temperature which correspond to the relativistic plasma in the early universe.

---
[1] Electronic address: masood@uhcl.edu

This paper is organized as follows: Section 2 briefly describes the calculational scheme which is used to compute electromagnetic properties of a QED medium and the parameters of the relativistic plasma in Section 3. Section 4 reports the results of sections 3 and discuss them in detail.

## 2. CALCULATIONAL SCHEME

The QED Lagrangian of such a system is written as

$$L = -\frac{1}{4} Z_3 F_{\alpha\mu\nu} F_\alpha^{\mu\nu} + i Z_2 \bar{\psi}_R \gamma_\mu D^\mu \psi_R + Z_2 \bar{\psi}_R m_0 \psi_R + e_0 Z_2 \sqrt{Z_3} \bar{\psi}_R \gamma_\mu A^\mu{}_R \psi_R \quad (1)$$

Renormalization techniques are not unique. More than one renormalizations schemes are available in statistical (hot and dense) medium also and have their own benefits and limitations. Mainly the real-time formalism [3] and the imaginary time formalism [4] are well-established. We use the covariant real-time formalism in Minkowski space which can easily check the KLN theorem at finite temperatures.

Interaction of propagating particles with the medium is incorporated by adding a statistical term to the photon propagator as:

$$\frac{1}{k^2} \rightarrow \frac{1}{k^2} - 2\pi i \delta(k^2) n_B(k),$$

and the photon distribution function is given as:

$$n_B(k) = \left[\frac{1}{e^{\beta k_0} + 1}\right]. \quad (1a)$$

β is the inverse temperature, i.e; 1/T. The corresponding fermion propagators is given as:

$$\frac{1}{p^2 - m^2} \rightarrow \left[\frac{1}{p^2 - m^2} + \Gamma_F(p)\right]$$

The fermion distribution function is expressed as

$$n_F(p) = \left[\frac{1}{e^{\beta(p_0)} + 1}\right] \quad (1b)$$

We can expand the distribution functions in a power series without any drastic approximations [5-8] as:

$$n_F \rightarrow \sum_{n=1}^{\infty} (-1)^n e^{-n\beta p_0} \quad (2)$$

It has been calculated that renormalization of QED in a hot medium of the early universe yields [4, 5] the self-mass (or self-energy) of electron as

$$\frac{\delta m}{m} \approx \frac{\alpha \pi T^2}{3m^2}\left[1 - \frac{6}{\pi^2} c(m\beta)\right] + \frac{2\alpha}{\pi} \frac{T}{m} a(m\beta) - \frac{3\alpha}{\pi} b(m\beta). \quad (3a)$$

With the wavefunction renormalization constant as

$$Z_2^{-1}(\beta) = Z_2^{-1}(T=0) - \frac{2\alpha}{\pi}\int_0^\infty \frac{dk}{k} n_B(k) - \frac{5\alpha}{\pi} b(m\beta)$$
$$+ \frac{\alpha T^2}{\pi v E^2} \ln\frac{1+v}{1-v}\left\{\frac{\pi^2}{6} + m\beta a(m\beta,\mu) - c(m\beta)\right\}, \qquad (2b)$$

The electron charge renormalization constant [7] is:

$$Z_3 \cong 1 + \frac{2e^2}{\pi^2}\left[\frac{ma(m\beta)}{\beta} - \frac{c(m\beta)}{\beta^2} + \frac{1}{4}\left(m^2 + \frac{1}{3}\omega^2\right)b(m\beta)\right]. \qquad (2c)$$

Which corresponds to the QED coupling constant because QED coupling α is related to the charge e through the relation α = e²/4π. The $a_i(m\beta)$ are given as:

$$a(m\beta) = \ln(1 + e^{-m\beta}), \qquad (3a)$$

$$b(m\beta) = \sum_{n=1}^{\infty}(-1)^n \operatorname{Ei}(-nm\beta), \qquad (3b)$$

$$c(m\beta) = \sum_{n=1}^{\infty}(-1)^n \frac{e^{-nm\beta}}{n^2}, \qquad (3c)$$

The calculation of $Z_3$, the charge renormalization is related to the vacuum polarization tensor due to the coupling of electromagnetic field with the medium of electromagnetically charged particles. Vacuum polarization tensor $\Pi_{\mu\nu}$ gives the polarization of electromagnetic waves in 4-dimensional space and explains the longitudinal component of polarization corresponding to the nonzero longitudinal component. $\Pi_{\mu\nu}$ plays a key role in computing the electromagnetic properties of the medium itself and leads to the calculation of plasma generating mass and then the Debye shielding length indicating the phase change in to the relativistic plasmas.

### 3. ELECTROMAGNETIC PROPERTIES and PLASMA SCREENING

It has been shown that the photons in this medium develop a plasma screening mass which can be obtained from the longitudinal and transverse component of the vacuum polarization tensor $\Pi_L(0,k)$ and $\Pi_T(k,k)$ where $K^2 = \omega^2 - k = 0$ in vacuum².

In this scheme of calculations, longitudinal and transverse components ($\Pi_L$ and $\Pi_T$, respectively) of vacuum polarization tensor $\Pi_{\mu\nu}$ play a crucial role in the calculation of the electromagnetic properties of a medium. The electromagnetic properties such as electric permittivity ε(K), magnetic permeability μ(K), refractive index $i_r$, propagation speed $v_{prop}$ and the magnetic moment $\mu_a$ of different particles in the medium are studied in the medium. Electric permittivity $\varepsilon(K)$ and the magnetic permeability $\mu(K)$ are related to $\Pi_L$ and $\Pi_T$ as [4, 8]:

$$\varepsilon(K) = 1 - \frac{\Pi_L}{K^2}, \qquad (4a)$$

$$\frac{1}{\mu(K)} = 1 + \frac{K^2 \Pi_T - \omega^2 \Pi_L}{k^2 K^2}, \qquad (4b)$$

And [16]

$$\varepsilon(K) = 1 + \chi_e \qquad (4c)$$

$$\mu(K) = 1 + \chi_m \qquad (4d)$$

Whereas, $\chi_e$ and $\chi_m$ give the dielectric constant and magnetization of a medium at a given temperature, respectively. The number density n can be evaluated as:

$$n(K) = \sqrt{\mu \varepsilon}$$

And the inverse of magnetic permeability correspond to the magnetic reluctance of the medium. The longitudinal and transverse components can be evaluated from the vacuum polarization tensor directly by using appropriate limits of photon frequency $\omega$ and the wavenumber k

$$\Pi_L \cong \frac{4e^2}{\pi^2}\left(1 - \frac{\omega^2}{\mathbf{k}^2}\right)\left[\left(1 - \frac{\omega}{2\mathbf{k}}\ln\frac{\omega+\mathbf{k}}{\omega-\mathbf{k}}\right)\left(\frac{ma(m\beta)}{\beta} - \frac{c(m\beta)}{\beta^2}\right)\right.$$
$$\left. + \frac{1}{4}\left(2m^2 - \omega^2 + \frac{11\mathbf{k}^2 + 37\omega^2}{72}\right)b(m\beta)\right], \qquad (5a)$$

and

$$\Pi_T \cong \frac{2e^2}{\pi^2}\left[\left\{\frac{\omega^2}{\mathbf{k}^2} + \left(1 - \frac{\omega^2}{\mathbf{k}^2}\right)\frac{\omega}{2\mathbf{k}}\ln\frac{\omega+\mathbf{k}}{\omega-\mathbf{k}}\right\}\left(\frac{ma(m\beta)}{\beta} - \frac{c(m\beta)}{\beta^2}\right)\right.$$
$$\left. + \frac{1}{8}\left(2m^2 + \omega^2 + \frac{107\omega^2 + 131\mathbf{k}^2}{72}\right)b(m\beta)\right]. \qquad (5b)$$

Substituting eqns. (5) in to eqns. (4), we get the dielectric constant of the medium as

$$\varepsilon(K) \cong 1 - \frac{4e^2}{\pi^2 K^2}\left(1 - \frac{\omega^2}{\mathbf{k}^2}\right)\left\{\left(1 - \frac{\omega}{2\mathbf{k}}\ln\frac{\omega+\mathbf{k}}{\omega-\mathbf{k}}\right)\left(\frac{ma(m\beta)}{\beta} - \frac{c(m\beta)}{\beta^2}\right)\right.$$
$$\left. + \frac{1}{4}\left(2m^2 - \omega^2 + \frac{11\mathbf{k}^2 + 37\omega^2}{72}\right)b(m\beta)\right\}, \qquad (6a)$$

and the magnetic reluctance $\frac{1}{\mu(K)}$ of the medium is

$$\frac{1}{\mu(K)} \cong 1 - \frac{2e^2}{\pi^2 k^2 K^2}\left[\omega^2\left\{\left(1 - \frac{\omega^2}{\mathbf{k}^4} - \left(1 + \frac{\mathbf{k}^2}{\omega^2}\right)\left(1 - \frac{\omega^2}{\mathbf{k}^2}\right)\frac{\omega}{2\mathbf{k}}\ln\frac{\omega+\mathbf{k}}{\omega-\mathbf{k}}\right)\right.\right.$$
$$\left.\left. \times\left(\frac{ma(m\beta)}{\beta} - \frac{c(m\beta)}{\beta^2}\right) - \frac{1}{8}\left(6m^2 - \omega^2 + \frac{129\omega^2 - 109k^2}{72}\right)b(m\beta)\right\}\right]. \qquad (6b)$$

Thermal contribution to electrical permittivity (Eq. 6a) and the magnetic permeability $\mu(K)$ (Eq. 6b) are used to evaluate the thermal contribution to the propagation speed of light and other relevant parameters in the early universe and it is given by

$$v_{prop} = \sqrt{\frac{1}{\varepsilon(K)\mu(K)}}. \qquad (7a)$$

which correspond to the refractive index $r_i$ of the medium as:

$$r_i = \frac{c}{v} = \sqrt{\frac{\varepsilon(K)\mu(K)}{\varepsilon_0(K)\mu_0(K)}} \qquad (7b)$$

The vacuum polarization tensor at finite temperature can be used to determine the phase of the medium indicating overall properties of the medium. Medium properties are changing with temperature due to modified electromagnetic couplings (2c). $K_L$ is evaluated by taking $\omega=k_0=0$ and p very small, the Debye shielding length $\lambda_D$ of such a medium is then given by the inverse of $K_L$

$$\lambda_D = 1/K_L \qquad (8a)$$

and the corresponding $\omega_D$ is

$$\omega_D = \frac{2\pi v_{prop}}{\lambda_D} = 2\pi K_L v_{prop} \qquad (8b)$$

Satisfying the relation

$$f_D \lambda_D = v_D \qquad (8c)$$

The plasma frequency $\omega_P$ is also related to $\omega_T$ from $\Pi_T(\omega = |k|)$ as

$$\omega_D = \frac{2\pi v_D}{\lambda_D} \qquad (9)$$

The longitudinal and transverse components of the photon frequency ω and the momentum k are evaluated in two different limits of magnitudes of ω and k as:

(I) ω= k₀= 0 in the limit of very low value of three momentum p is very small.

$$K_L^2 = \Pi_L(0,k) \cong \frac{4e^2}{\pi^2}\left[\left(\frac{ma(m\beta)}{\beta} - \frac{c(m\beta)}{\beta^2}\right) + \frac{1}{4}\left(2m^2 + \frac{11\mathbf{k}^2}{72}\right)b(m\beta)\right], \qquad (10a)$$

$$K_T^2 = \Pi_T(0,k) \cong \frac{2e^2}{\pi^2}\left[\frac{1}{8}\left(2m^2 + \frac{131\mathbf{k}^2}{72}\right)b(m\beta)\right]. \qquad (10b)$$

and the magnitude of vector K is calculated as

$$|K| = (K_L^2 + K_T^2)^{1/2} \qquad (10c)$$

(II)     $\omega=|p|$ and the magnitude of 3 momentum p is still very small

$$\omega_L^2 = \Pi_L(|k|,k) \cong 0 \qquad (11a)$$

$$\omega_T^2 = \Pi_T(|k|,k) \cong \frac{2e^2}{\pi^2}\left[\frac{ma(m\beta)}{\beta} - \frac{c(m\beta)}{\beta^2} + \frac{1}{8}\left(2m^2 + k^2 + \frac{238\,\mathbf{k}^2}{72}\right)b(m\beta)\right]. \qquad (11b)$$

Such that

$$\omega = (\omega_L^2 + \omega_T^2)^{1/2} \qquad (11c)$$

The plasma frequency is defined as $\omega_p^2 = \omega_T^2$ as $\omega_L^2 = 0$ and the Debye shielding length is obtained by equation (10b). These results can easily be generalized to different situations using the initial values of $\Pi_L$ and $\Pi_L$ from equations (5).

## 4. RESULTS and DISCUSSIONS

The physically measureable parameters of QED in certain ranges of photon frequency and momentum are given in Eqns. (8-11). We have studied the temperature dependence of QED parameters as functions of temperature, measured in units of electron mass. Natural system of units is used such that c=1 and mass, energy and momentum are expressed in units of MeV. All of the QED parameters such as electric permittivity, magnetic permeability, magnetic reluctance, dielectric constant, refractive index and the propagation speed are also normalized to unity in vacuum to compare with the corresponding thermal corrections up to the one loop level. Plots of these functions help to understand the behavior of QED parameters with the increase in temperature. It also shows how the plasma screening was dissolved near the nucleosynthesis temperature (T ≈ m) when electrons and photons were not shielding each other and nucleosynthesis started due to the electron-neutrino interactions.

Figure (1) shows the plot of $k_L^2$ and $\omega_T^2$ as a function of temperature. These two functions have quadratic dependence on temperature though the change in frequency is slower than the momentum. Therefore the change in longitudinal component of propagation vector is faster than the transverse frequency showing the fast bending but slower increase in transverse oscillation.

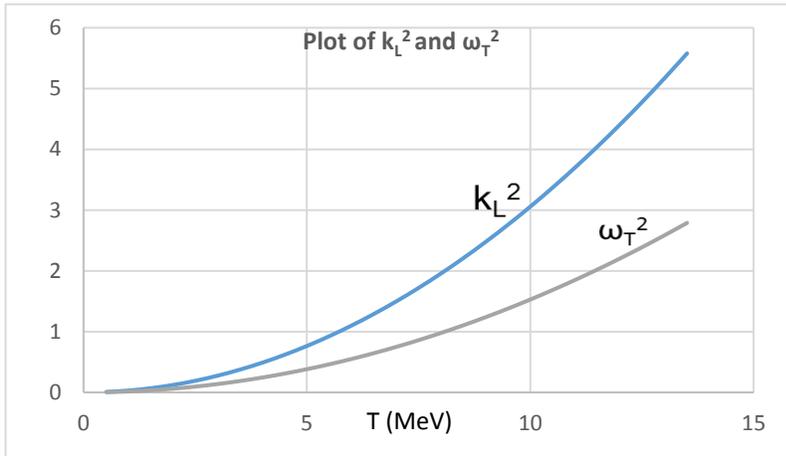

*Figure 1: Temperature dependence of the longitudinal component of the propagation vector (square) and the transverse frequency (square) are plotted showing that the Debye shielding length is greater than the plasma frequency.*

On the other hand, it is obvious from Figure (2) that the transverse component of propagation vector is not affected significantly as it just depends on the b(mβ) only which gives ignorable contribution as compared to quadratic function c(mβ). Moreover, we know that $k_L^2=0$ at low temperature and it just appears to contribute at $T>10^9K$ (temperature corresponding to the electron mass). Negligible change in $k_T^2$ gives the assurance that the speed remains almost the same thus the $k_L^2$ has quadratic dependence on T which indirectly affect the propagation velocity by mainly bending and then slowing it in the plasma.

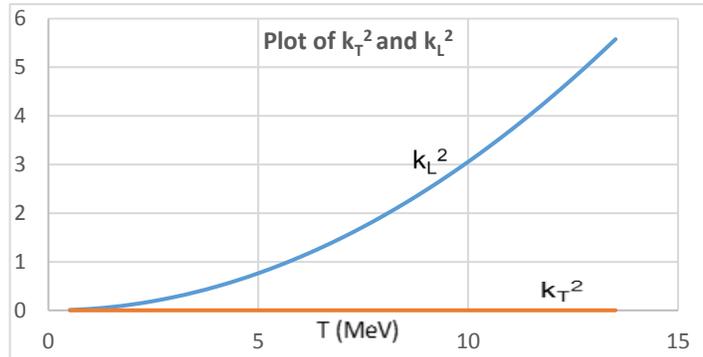

*Figure 2: Comparing the plots of $k_L$ (square) and $k_T$ (square), it can be clearly seen that propagation vector in the transverse direction is independent of Temperature. Temperature increases the inverse of Debye length ($1/k_L$) but transverse motion will not be affected*.

When the plasma screening is large enough, it traps the light and the speed of light is not a measurable speed any more. Propagation with the transverse frequency makes it phase velocity which is responsible for oscillation of light waves within the Debye sphere of length proportional to $1/k_L$. To clearly identify the unique behavior of light in this range by comparison, we plot the magnitude of wave vector k along with $k_L^2$ and $\omega_T^2$ in Figure 3.

It is obvious that k is directly proportional to T and is contributed both by the longitudinal and transverse components. When the temperature reaches around 12 MeV or higher the contribution of the transverse frequency dominates over the momentum. On the other hand the longitudinal component of the wavenumber is an inverse of the Debye length, increasing $k_L$ indicates the decrease in the Debye shielding length with temperature. Quadratic increase in $k_L^2$ correspond to inverse proportionality of Debye length with temperature.

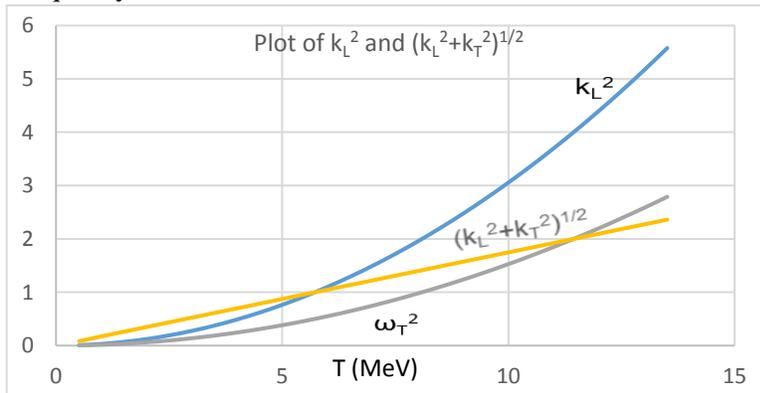

*Figure 3: Comparing the plots of k, the magnitude of the propagation vector, longitudinal propagation $k_L$ (square) and transverse frequency $\omega_T$ (square).*

It is also clearly seen in the graph that the transverse component of k is significant at T< 6MeV. Then the transverse contribution is reduced as compared to the longitudinal component.

A plot of plasma parameters in Figure 4 shows that the Debye length decreases with increase in plasma frequency at rising temperature. It means that the system behaves as plasma for the region of temperature 1< T/m <16. Therefore, the proper QED plasma phase in the universe was there only when T < 16 MeV.

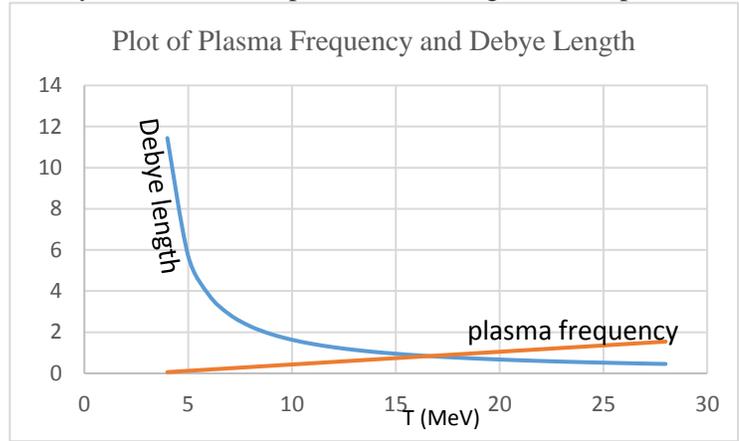

*Figure 4: Debye length ($1/k_L$) and the Plasma frequency are plotted as a function of temperature showing that the screening length increases with T. When temperature rises around 16 times the mass of electron which is around 8MeV, a phase transition takes place and the plasma frequency becomes greater than the shielding length.*

A phase transition occurred at around 16 MeV and an unusual behavior of trapping of light was shown above that temperature. It also explains the extremely large coupling constant of QED.

We have plotted some of the important QED parameters as functions of temperature including the electron mass, charge, dielectric constant, magnetic reluctance and the propagation speed of the medium. It can be easily seen that the electric charge and the propagation speed are slowly changing function of temperature and are almost independent of temperature below 5MeV. However, mass of electron, dielectric constant and magnetic reluctance of the medium changes significantly and their quadratic behavior is obvious.

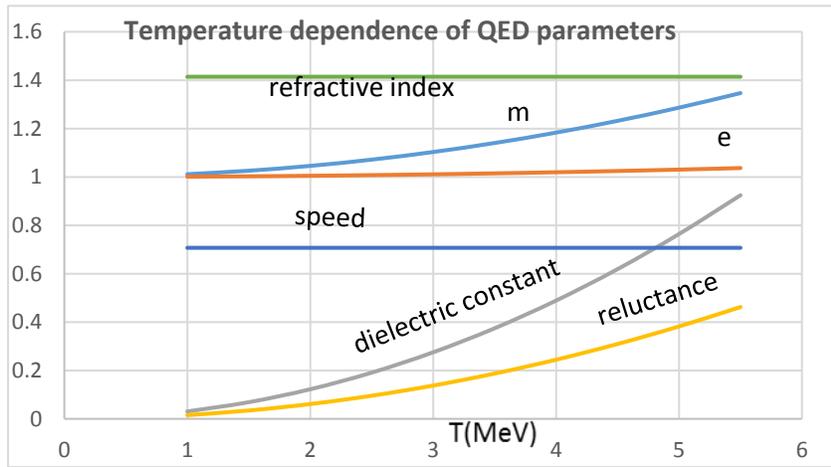

*Figure 5: All of the plasma parameters are plotted as functions of temperature such that the reluctance, propagation speed, -dielectric constant are all increasing with temperature.*

Thus the electromagnetic properties of such a medium are tremendously changed with temperatures though the light propagation is not much affected due to the reason that the transverse properties are almost same whereas the medium mainly affects the longitudinal behavior of light and a plasma screening is there in the longitudinal plane only.

Eqns.(7-11) in section 3 give the properties of the medium where the plasma effects can be seen. However, the photon propagation through the medium could be studied for other relevant ranges of the photon frequency and wavenumber also.

(i) $\omega \gg k$

$$\pi_L = -\frac{\omega^2 e^2 T^2}{3k^2} \qquad (12a)$$

$$\pi_T = \frac{\omega^2 e^2 T^2}{6k^2} \quad (12b)$$

$$\varepsilon(k) = 1 + \frac{\omega^2 e^2 T^2}{3k^4} \quad (13a)$$

$$\frac{1}{\mu(k)} \cong \frac{\omega^4 e^2 T^2}{3k^4 K^2} \quad (13b)$$

$$v_{prop} = \sqrt{\frac{\omega^4 e^2 T^2}{K^2(3k^4 + \omega^2 e^2 T^2)}} \quad (14a)$$

$$r_i = \frac{c}{v} = c\sqrt{\frac{K^2(3k^4 + \omega^2 e^2 T^2)}{\omega^4 e^2 T^2}} \quad (14b)$$

Taking $\omega^2 \approx T^2$ at extremely high temperatures, we obtain the effective parameters of the theory as

$$m_{eff} \approx m(1 + \frac{\delta m}{m}) \approx m(1 + \frac{\alpha \pi T^2}{2m^2}) \quad (15a)$$

$$\alpha_{eff} \approx \alpha(1 + \delta\alpha/\alpha) \approx \alpha(1 + \frac{\alpha T^2}{6m^2}) \quad (15b)$$

$$\pi_L = -\frac{e^2 T^2}{3k^2} \qquad \pi_T = \frac{e^2 T^2}{6k^2} \quad (16)$$

$$\varepsilon(k) = 1 + \frac{e^2 T^4}{3k^4}; \qquad \frac{1}{\mu(k)} \cong 1 + \frac{e^2 T^4}{3k^4} \quad (17)$$

$$v_{prop} = \sqrt{\frac{e^2 T^4}{(3k^4 + e^2 T^4)}}. \quad (18)$$

$$r_i = \frac{c}{v} = \sqrt{\frac{(3k^4 + e^2 T^4)}{e^2 T^4}} \approx 1 \quad (19)$$

For small values of k, the vacuum relation $v_{prop} \approx c = 1$ is still valid and $v_{prop} \approx 1 \approx r_i$ is reproduced or such a medium can be considered as a transparent medium. Small k value indicates that the wave number is negligible and the photon frequency is very high. It means that the medium of the early universe was transparent for the extremely high energy radiation which is a naturally acceptable result. We expect the extremely high energy radiation in the early universe and that is what we should expect in the early universe. Understanding of propagation of light will help us to understand the inflation and generation of anisotropy [17-20] in the early universe. A similar type of effect has already been seen in QCD using the perturbative calculations of QCD in the real-time formalism [21].